\documentclass[a4paper]{spie}  
\usepackage{subcaption}
\usepackage{float}
\usepackage{amsmath,amsfonts,amssymb}
\usepackage{graphicx}
\usepackage[colorlinks=true, allcolors=blue]{hyperref}

\newcommand{\apj}{\textit{The Astrophysical Journal}}
\newcommand{\aap}{\textit{Astronomy \& Astrophysics (A\&A)}}
\newcommand{\solphys}{\textit{Solar Physics}}
\newcommand{\aj}{\textit{Astronomical Journal}}

\title{CUbesat Solar Polarimeter (CUSP) Sensitivity Estimation and Performance Optimization using
Geant4}

\author[a]{Abhay~Kumar}
\author[a,c]{Giovanni~Lombardi}
\author[g]{Giovanni~De~Cesare}
\author[a]{Nicolas~De~Angelis}
\author[a]{Sergio~Fabiani}
\author[a]{Ettore~Del~Monte}
\author[a,b]{Andrea~Alimenti}
\author[g,m]{Riccardo~Campana}
\author[a]{Enrico~Costa}
\author[a]{Paolo~Soffitta}
\author[h]{Mauro~Centrone}
\author[a]{Sergio~Di~Cosimo}
\author[a]{Giuseppe~Di~Persio}
\author[a]{Alessandro~Lacerenza}
\author[a]{Pasqualino~Loffredo}
\author[a]{Fabio~Muleri}
\author[m]{Paolo~Romano}
\author[a]{Alda~Rubini}
\author[a]{Emanuele~Scalise}
\author[a,b]{Enrico~Silva}
\author[d]{Davide~Albanesi}
\author[e]{Ilaria~Baffo}
\author[f]{Daniele~Brienza}
\author[d]{Valerio~Campamaggiore}
\author[i]{Giovanni~Cucinella}
\author[j]{Andrea~Curatolo}
\author[d]{Giulia~de~Iulis}
\author[d]{Andrea~Del~Re}
\author[i]{Vito~Di~Bari}
\author[i]{Simone~Di~Filippo}
\author[f]{Immacolata~Donnarumma}
\author[e]{Pierluigi~Fanelli}
\author[k]{Nicolas~Gagliardi}
\author[d]{Paolo~Leonetti}
\author[f]{Matteo~Mergè}
\author[l]{Gabriele~Minervini}
\author[j,k]{Dario~Modenini}
\author[i]{Andrea~Negri}
\author[j]{Daniele~Pecorella}
\author[i]{Massimo~Perelli}
\author[k]{Alice~Ponti}
\author[d]{Francesca~Sbop}
\author[j,k]{Paolo~Tortora}
\author[f]{Alessandro~Turchi}
\author[f]{Valerio~Vagelli}
\author[f]{Emanuele~Zaccagnino}
\author[d]{Alessandro~Zambardi}
\author[e]{Costantino~Zazza}

\affil[a]{INAF-IAPS, Via del Fosso del Cavaliere 100, 00133 Rome, Italy}
\affil[b]{Department of Industrial, Electronic and Mechanical Engineering, Roma Tre University, Via V. Volterra 62, 00146, Roma, Italy}
\affil[c]{Department of Enterprise Engineering ”Mario Lucenti”, University of Rome Tor Vergata, Via Cracovia 50, 00133 Rome, Italy}
\affil[d]{DEDA Connect s.r.l., Via Vincenzo Lamaro 51, 00173 Rome, Italy}
\affil[e]{DEIM, University of Tuscia, Largo dell’Università, 01100 Viterbo, Italy}
\affil[f]{ASI, Via del Politecnico snc, 00133, Roma, Italy}
\affil[g]{INAF-OAS Bologna, Via Gobetti 93/3, 40129, Bologna, Italy}
\affil[h]{INAF-OAR, Via Frascati 33, 00040, Monte Porzio Catone, Italy}
\affil[i]{IMT s.r.l., via Carlo Bartolomeo Piazza 30, 00161 Rome, Italy}
\affil[j]{Department of Industrial Engineering - Alma Mater Studiorum University of Bologna - Via Montaspro 97, 47121 Forlì, Italy}
\affil[k]{Interdepartmental Centre for Industrial Aerospace Research - Alma Mater Studiorum University of Bologna - Via Carnaccini 12, 47121 Forlì, Italy}
\affil[l]{INAF-Headquarters, Viale del Parco Mellini 84, 00136, Roma, Italy}
\affil[m]{INAF-OACT, Via S. Sofia 78, 95123, Catania, Italy}
\affil[n]{INFN Bologna Section, viale Berti Pichat 6/2, 40127, Bologna, Italy}

\authorinfo{Further author information: (Send correspondence to Abhay Kumar)\\Abhay Kumar: E-mail: abhay.kumar@inaf.it, Telephone:  +39 06 4993 4098}

\begin{document} 
\maketitle

\begin{abstract}
The CUbesat Solar Polarimeter (CUSP) aims to measure the linear polarization of solar flares in the 25 – 100~keV X-ray band using a Compton scattering polarimeter. CUSP will allow us to study the magnetic reconnection and particle acceleration in the flaring magnetic structures of our star by providing high-sensitivity polarization measurements. CUSP is a project in the framework of the Alcor Program\footnote{\url{https://www.asi.it/en/technologies-and-engineering/micro-and-nanosatellites/alcor-program/}, consulted on 12 Jul 2025} of the Italian Space Agency aimed to develop innovative CubeSat technologies and missions.
As part of CUSP's Phase B study, which began in December 2024
and will continue for one year, we present the development status of the Geant4 based simulator to accurately simulate the detector’s response and initial results on the sensitivity of the instrument. Geant4 Monte Carlo simulation is used to assess the physical interactions of the source photons with the detector and the passive materials. We implemented a detailed CUSP Mass Model within Geant4 to simulate and estimate the instrument's sensitivity, correcting the geometric effects of the instrument. We also evaluated the effect of backscattering shielding on the sensitivity to optimize the mass model of the instrument.  
\end{abstract}


\section{INTRODUCTION}
\label{sec:intro}  

Sun is the nearest star and primary energy source for life on Earth. The violent, energetic phenomenon by which Sun releases energy is called Solar flares (SFs). In SFs, a substantial amount of energy is released due to magnetic reconnection, and particles are accelerated towards the interplanetary space and lower layers of the solar atmosphere along the magnetic field lines. It can affect space weather and human activities on the ground. Despite over a century of studying the Sun, the physical process responsible for the magnetic field of the Sun and the heating of the outer atmosphere of the Sun called Corona ( $>$one million Kelvin), having a temperature more than the solar photosphere (6000 K) \cite{klimchuk2006,moortel2015} is not fully understood. It is now widely accepted that the magnetic field lines protruding from the solar photosphere play a crucial role in heating the Corona. The solar X-ray spectrum mainly consists of emission lines at low energy below 10~keV and thermal and non-thermal components due to Bremsstrahlung above about 10~keV. Theoretical models predict distinct polarization signatures for thermal and non-thermal emission mechanisms \cite{emslie1980,zharkov2010}, which may appear degenerate when interpreted solely through spectroscopy. Additionally, the degree and orientation of X-ray polarization also depends on factors such as particle beaming, magnetic field geometry, and the observer’s viewing angle \cite{Jeffrey2020}. Therefore, polarization measurements are essential for breaking degeneracies between competing models. In particular, during the early impulsive phase of a solar flare, the X-ray energy spectrum is dominated by non-thermal bremsstrahlung emission from accelerated electrons. As the flare evolves, plasma heating increases the contribution of thermal bremsstrahlung, which typically overtakes the non-thermal component in the 10–30 keV energy range, thereby reducing the overall polarization degree \cite{Nagasawa_2022,Grigis2004,dennis2005}. This necessitates studying solar flares with sufficient temporal resolution. 

Despite its diagnostic potential, X-ray polarization measurements of solar flares have been few and statistically limited, generally yielding only upper limits or marginal detections to date \cite{Tindo1970,Tindo1972a,Tindo1972b,Tramiel1984,suarez2006,boggs06}. Advancing this field necessitates the development of dedicated instruments with high sensitivity and sufficient temporal resolution to capture the impulsive and rapidly evolving nature of flare events. To address this, the CUSP instrument is designed to perform time-resolved polarization measurements in the 25-100~keV energy band. This capability will allow us to understand the flare emission mechanisms and the associated magnetic and particle acceleration processes. CUSP has been proposed as a constellation of two 6U-XL CubeSats consisting of one orbital plane and orbiting with a time variable phase difference to maximise the observation time of the Sun, and will proceed with a single-satellite asset in its baseline implementation.

The activity presented is aimed to develop a Geant4 based simulator and analysis method to understand the polarimetric response of the CUSP and optimise the payload design. We introduce the polarisation analysis method based on the Stokes parameter formalism and initial results on the polarimetric sensitivity of the instrument. The assessment of the spurious modulation, effective area, and modulation factor at 60~keV using polarised and unpolarised simulations for edge and corner angles is carried out. The backscattered photons from passive material are not polarised and can induce a systematic effect on the modulation curve. The effect of molybdenum shielding to mitigate the backscattering from the passive material of the detector is also discussed. This assessment is needed before starting a comprehensive study of the CUSP polarimeter performance, including a refinement of the polarimeter sensitivity (Minimum Detectable Polarisation-MDP) \cite{sergio2022}.  

\section{Principles and methodology}

\subsection{Principle of X-ray polarisation measurements}
\label{sec:principle}

\begin{figure}[H]
  \centering
  \includegraphics[trim=0pt 20pt 0pt 30pt, clip,scale=0.7]{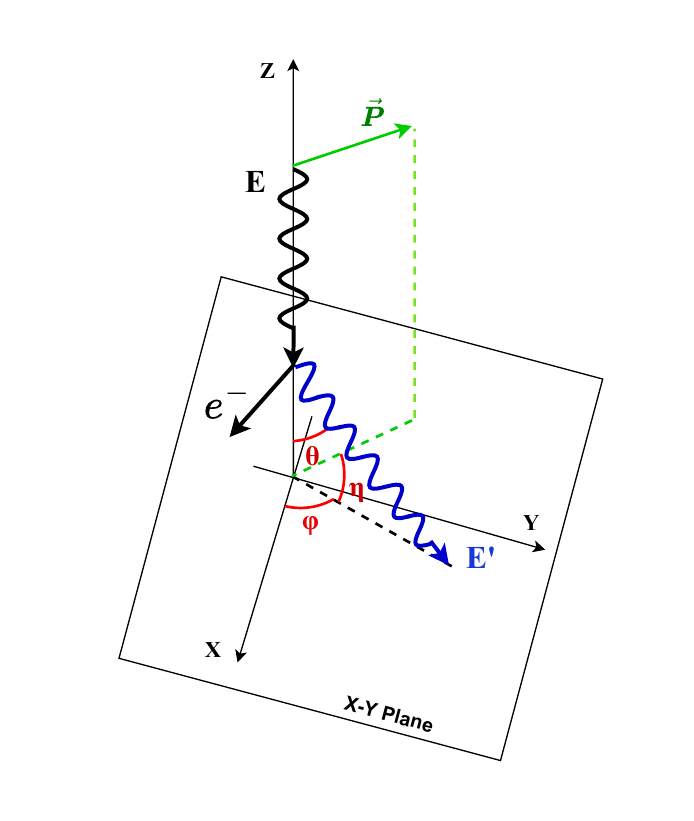} 
  \caption{Schematic of Compton scattering of a linearly
polarized photon}
  \label{fig:compt_scatt}
\end{figure}

In the CUSP energy range, Compton scattering is the dominant process of photon - matter interaction. 
The Compton scattering is sensitive to the polarization of the scattered photon because the scattering direction of the photon is affected by the incident photon's electric field. Polarised radiation induces a
preferential azimuthal angular direction of scattering (normal to the incident beam axis) as described by the Klein-Nishina cross section, which gives the differential cross section of the process under the simple hypothesis of scattering on a free electron at rest \cite{heitler54}:

$$\frac{\mathrm{d}\sigma}{\mathrm{d}\Omega}=\frac{r_{0}^{2}}2\left(\frac{E^{\prime}}E\right)^{2}\left(\frac E E+\frac{E^{\prime}}E-2\sin^{2}\theta\cos^{2}\eta\right)$$

where,
$${\frac{E^{\prime}}{E}}=\left[1+{\frac{E}{m_{\mathrm{e}}c^{2}}}(1-\cos\theta)\right]^{-1}$$

where $r_\circ$ is the classical electron radius, m$_{e}$c${^2}$ denotes the rest mass energy of an electron, E and E$^{'}$ are the incident and scattered  photon energy, respectively, $\theta$ is the Compton scattering angle, and $\eta$ is the angle forms between
the plane of scattered photon direction with the plane containing the polarization direction of the incident photon as shown in Figure \ref{fig:compt_scatt}


The azimuthal scattering direction is modulated as $\cos^{2}\eta$ and peaks in the direction orthogonal to the polarisation of incident photons (scattering cross section is maximum for $\eta=\pi/2$), implying linearly polarised photons are preferentially scattered orthogonal to their polarisation direction. The polarisation sensitivity of the polarimeter is defined by the modulation factor $\mu(\theta)$, which is the fraction of modulated signal corresponding to 100$\%$ polarised radiation. It is given by

$$\mu(\theta)={\frac{\sin^{2}\theta}{{\frac{E^{\prime}}{E}}+{\frac{E}{E^{\prime}}}-\sin^{2}\theta}}$$

For a 100 $\%$ linearly polarized photon beam, the modulation factor is called $\mu_{100}$. The polarisation angle and degree of polarisation are retrieved from the azimuthal distribution of scattering angle $\varphi$, which is the angle between the polarized scattered photon plane and the reference axis (X-axis here), as shown in Figure \ref{fig:compt_scatt}. Such an azimuthal scattering angle distribution is supposed to retain the same cosine-squared modulation as the differential cross-section. The amplitude of this cosine-squared distribution gives the modulation factor, and the phase $\pm$90 gives the polarisation angle.

\subsection{Polarimeter configuration}

The CUSP payload is a dual-phase Compton scattering polarimeter \cite{Costa1995} optimized for the 25–100~keV energy range. It employs coincident detection of Compton-scattered photons using a modular assembly of 64 low-Z plastic scintillator bars (scatterers) and 32 high-Z GAGG:Ce ($\mathrm{Gd}_3\mathrm{Al}_2\mathrm{Ga}_3\mathrm{O}_{12}$:Ce) scintillator bars (absorbers). The low-Z plastic bars maximise the probability of Compton scattering, while the GAGG crystals efficiently absorb the scattered photons via photoelectric interaction. The azimuthal distribution of these coincident events encodes the polarization information. All scintillator elements are individually wrapped in reflective material to confine scintillation light and ensure optical isolation, reducing crosstalk and preserving signal quality.

\begin{figure}[H]
  \centering
  \includegraphics[trim=160pt 0pt 200pt 20pt, clip,scale=0.8]{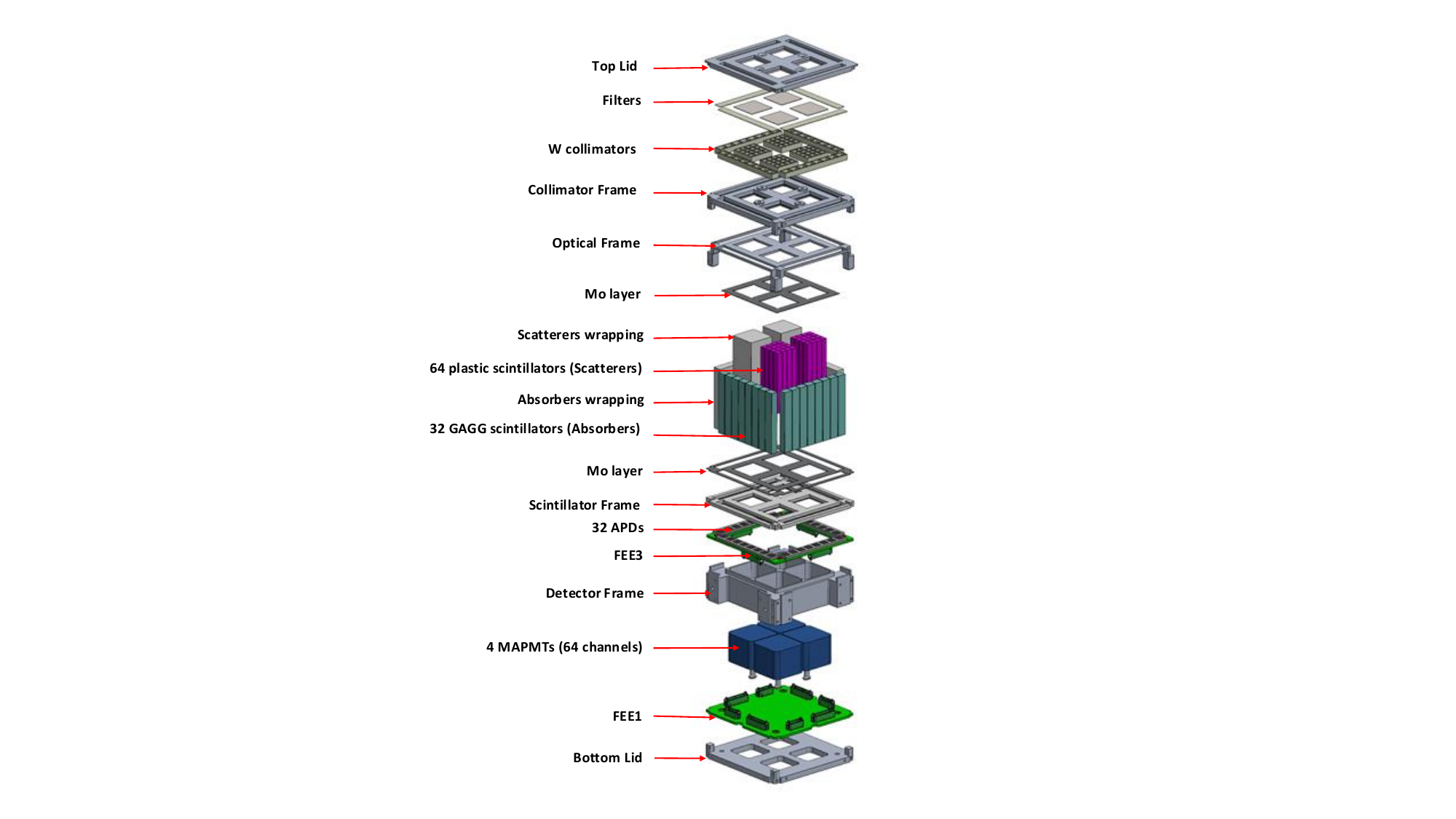} 
  \caption{The scheme of the CUSP polarimeter}
  \label{fig:cusp_design}
\end{figure}

The plastic bars are grouped into four 4$\times$4 matrices, each coupled to a 16-anode Multianode Photo-Multiplier Tube (MAPMT), offering high quantum efficiency and fast timing. The surrounding GAGG bars are arranged in a square configuration into four linear 1$\times$8 arrays, each read out by Avalanche photo diodes (APDs). Each scintillator bar is fitted with a tungsten collimator and X-ray filter to limit its field of view to approximately $\pm$36$^{\circ}$ and effectively suppressing the background to improve detection sensitivity. The exploded view of the detector assembly is shown in Figure \ref{fig:cusp_design}.

\subsection{Geant4 simulation framework and analysis method}\label{sec:method}

A dedicated Monte Carlo simulation framework written in C++ and based on Geant4 \cite{agostinelli03_short} is used to model the physical interactions of hard X-ray photons with both the active detector elements and surrounding passive materials. A detailed mass model of the CUSP polarimeter is implemented by converting a CAD model into Geometry Description Markup Language (GDML) format, which is then directly imported into the Geant4 detector geometry construction. The material properties of different GDML model components are defined in the detector construction file. Additionally, the particle source configuration, beam size, physics list, and data output are defined. The monochromatic polarised and unpolarised source is used to simulate the polarimetric response of the polarimeter. The realistic geometric model used in this work corresponds to the mechanical configuration described in the previous section.

\begin{figure}[H]
  \centering
  \includegraphics[trim=0pt 0pt 0pt 0pt, clip,scale=0.25]{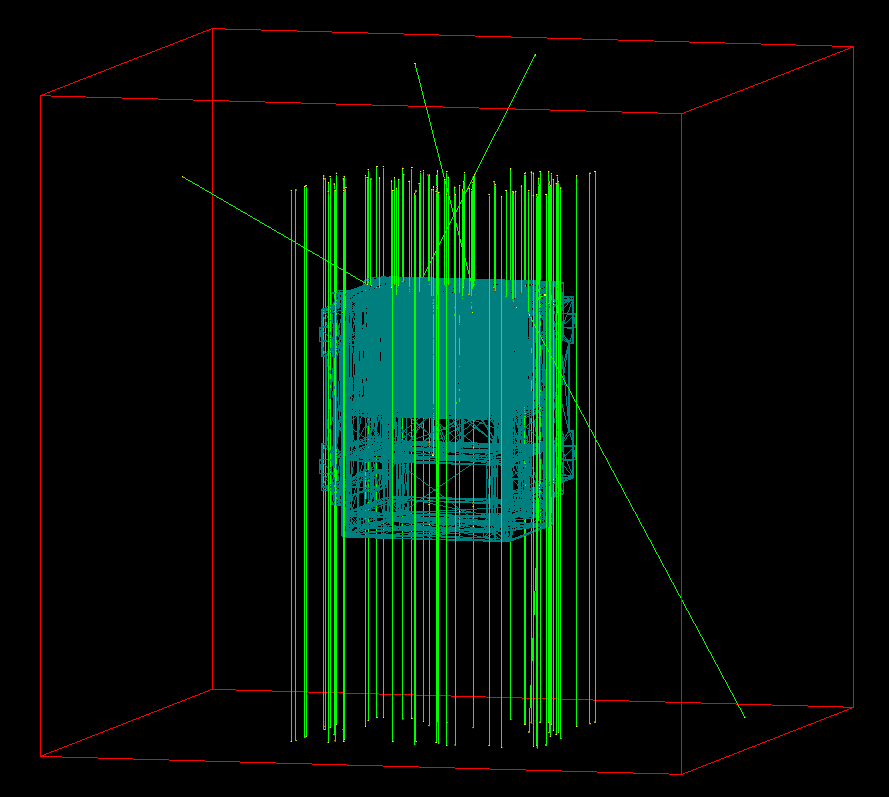} 
  \caption{The parallel beam of 60~keV photons (green lines) impinging on the CUSP polarimeter inside the world volume (red cube) in Geant4}
  \label{fig:geant4_sim}
\end{figure}

The simulation framework supports both interactive and batch execution modes. The interactive mode allows real-time visualization of the detector geometry and photon interaction processes, allows debugging, validation, and helps in optimization of the mass model. Figure \ref{fig:geant4_sim} shows the photons interacting with the sensitive and passive volume of the CUSP visualised in interactive mode.  The batch mode enables large-scale event simulations, producing output files containing detailed information on photon interactions. In batch mode, the X-ray source and its properties are defined in the macro file using the General Particle Source (GPS) module in Geant4. For each incident photon, the output includes a list of sensitive detector volume name where energy is deposited, interaction position (center of bar as default), deposited energy, and the information about the incident photon, which is then used for subsequent data analysis.

\begin{figure}[H]
  \centering
  \includegraphics[trim=0pt 0pt 0pt 30pt, clip,scale=0.45]{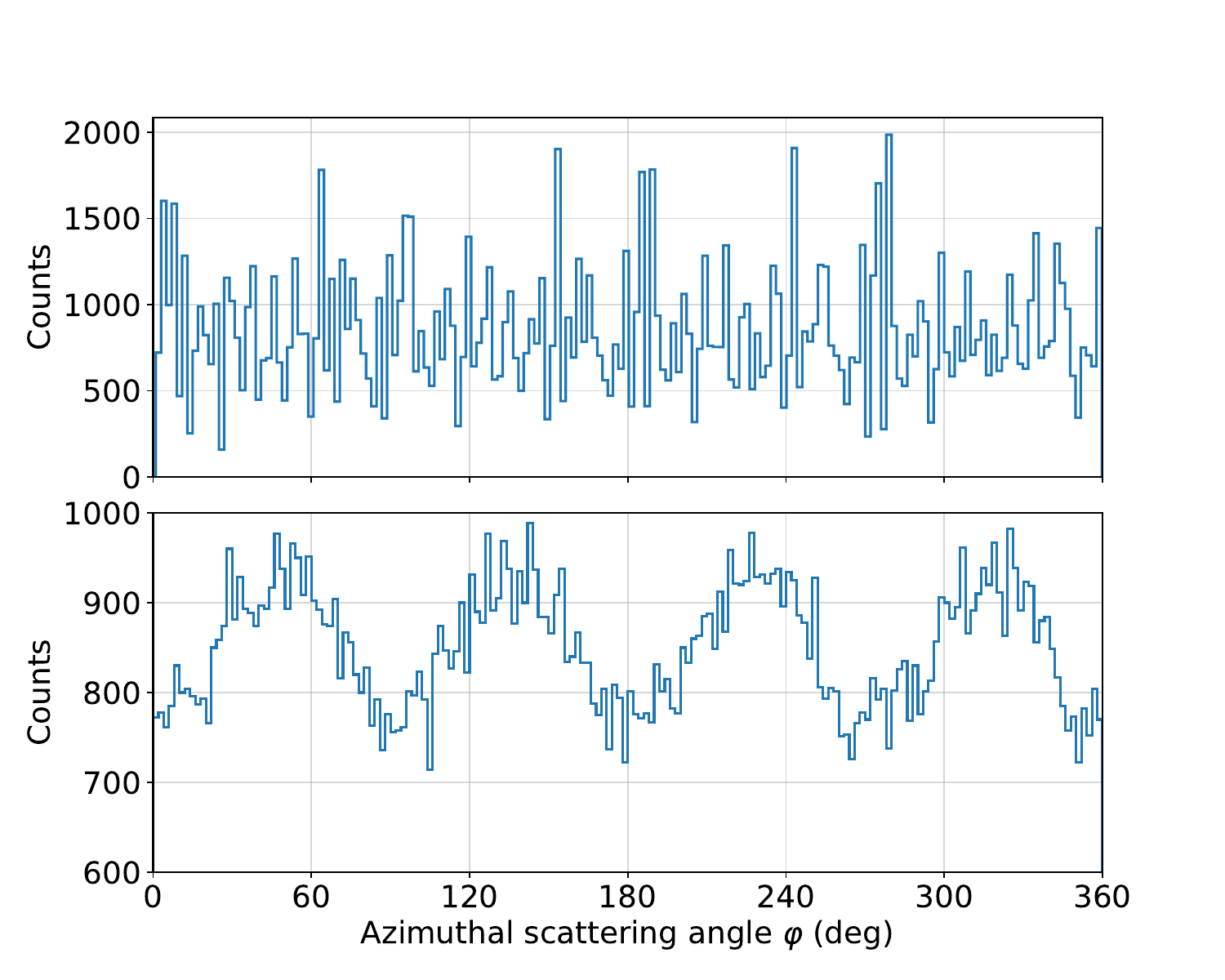} 
  \caption{\textbf{Top panel}: the modulation curve obtained by assigning to each coincident event an azimuthal angle of scattering derived by the positions of the interacting bars (center of each bar footprint). \textbf{Bottom panel}: the randomization is applied. The typical four peaks of the spurious modulation patter of a squared geometry emerges.}
  \label{fig:randomised_spurious}
\end{figure}

For the data analysis, single coincident events are filtered, having one hit in the scatterer and one in the absorber, to extract the azimuthal scattering angle. It is estimated using the interaction position information in the scatterer and absorber. The photon interaction via Compton scattering (in plastic) and then the photoelectric effect (in GaGG) can occur anywhere in the bars. A bar position is identified by the center of the footprint in the polarimeter design in simulation (considering a realistic scenario), and the actual photon interaction point is derived by randomizing the interaction point in the bar. This randomization is performed in the entire bar volume in case of the plastics and in a 1 mm (based on attenuation length in GAGG) thin layer of the GaGG facing the plastics bars. CUSP does not detect the photon interaction position along the bar elongation, thus such a coordinate is not randomized. When simulating an unpolarized beam of radiation, such randomisation enables the retrieval of the spurious modulation pattern. For a square detector geometry, the characteristic four-peak structure emerges after randomisation (see Figures \ref{fig:randomised_spurious} and \ref{fig:mod_curve}). The square geometry of the CUSP (see Figure \ref{fig:cusp_design}) inherently introduces modulation due to the geometry, even for the unpolarised radiation, which is called spurious modulation. The spurious modulation needs to be decoupled from the signal to extract the true source modulation and accurately determine the polarisation degree and angle. 

The estimation of polarisation degree and angle can be carried out using different techniques proposed in the literature \cite{DiMarco2022,sk13,kislat2015}. This assessment is carried out by exploring the technique of Strokes parameters estimation based on an unbinned event-by-event approach as done for the IXPE data analysis  \cite{DiMarco2022,rankin2022}. The Stokes parameters are additive in nature and allow direct correction for the spurious effect at a single event level. The computation of the Stokes parameters corresponds to using the second harmonic in the signal (the cos$^{2}$ ) and only the contribution of the spurious modulation to it is relevant.

For each detected coincident event, we compute the set of stokes parameters

$$\begin{array}{c}{{q_{i}=2\cos\left(2\phi_{i}\right)}}\\ {{u_{i}=2\sin\left(2\phi_{i}\right)}}\end{array}$$

 Where $\phi_{i}$ is the most likely azimuthal angle of the electric field vector \cite{kislat2015}. In case of scattering polarimeters, $\phi_{i} = \varphi_{i}+90^{\circ}$ as photon preferentially scatter perpendicular to its electric field direction as discussed in section \ref{sec:principle} where azimuthal scattering angle  $\varphi_{i}$ is calculated after randomizing the interaction bar positions as discussed above.

For the observation of N events, the normalised Stokes parameters q and u can be computed: 
$${{q={\frac{\sum_{i}q_{i}}{N}}}}, \qquad  {{u={\frac{\sum_{i}u_{i}}{N}}}}$$

with associated uncertainties given by 
$${{\sigma_{q}=\sqrt{\frac{2-q^{2}}{N-1}}}}, \qquad {{\sigma_{u}=\sqrt{\frac{2-u^{2}}{N-1}}\,.}}$$

From the q and u modulation factor can be estimated:
$$m={\sqrt{q^{2}+u^{2}}}$$
 
The polarisation degree and angle are given by

$$p=\frac{m}{\mu_{100}}, \qquad \varphi={\frac{1}{2}}\tan^{-1}\left({\frac{u}{q}}\right)$$

The uncertainties on these for measurements of high statistical significance, can be estimated as:
$$\sigma_{m}\approx{\sqrt{\frac{2-m^{2}}{N-1}}}, \qquad \sigma_{\varphi}\approx{\frac{1}{m{\sqrt{2(N-1)}}}} \qquad \mathrm{for}\ m/\sigma_{m}\gg1$$

A polarimeter can have systematics that lead to the presence of a modulation due to the instrument, even for the unpolarised source. This can be calibrated and subtracted to get the correct modulation factor and angle.
There are two ways (M1 and M2) of disentangle the spurious component from the source described in Rankin et al.\cite{rankin2022} to get the Stokes parameters of source ($q_{\mathrm{source}}$, $u_{\mathrm{source}}$) and spurious component ($q_{\mathrm{sm}}$, $u_{\mathrm{sm}}$). This technique is already tested for IXPE data in flight and during ground calibration \cite{rankin2022,DiMarco2022,Rankin2023spie}:

M1) Subtracting the Stokes parameters measured with unpolarised simulations from those of polarised ones ($q_{\mathrm{meas}}$, $u_{\mathrm{meas}}$). 
$${{q_{\mathrm{source}}=q_{\mathrm{meas}}-q_{\mathrm{sm}}}}$$
$${{u_{\mathrm{source}}=u_{\mathrm{meas}}-u_{\mathrm{sm}}}}$$

M2) Subtracting the Stokes parameters for the two polarised simulations with polarisation angles 90 degrees apart.
$$q_{\mathrm{soure}}={\frac{q_{0}-q_{90}}{2}}$$
$$u_{\mathrm{soure}}={\frac{u_{0}-u_{90}}{2}}$$

The spurious modulation can be estimated in this case by adding the two polarised simulations with polarisation angles 90 degrees apart.
$${{q_{\mathrm{sm}}={\displaystyle\frac{q_{0}+q_{90}}{2}}}}$$
$$ {{u_{\mathrm{sm}}={\displaystyle\frac{u_{0}+u_{90}}{2}}.}}$$

\section{Simulation results and performance evaluation}

The estimation of the modulation factor and the assessment of the molybdenum shielding to mitigate the backscattered photons from passive materials around the sensitive volume of the polarimeter are carried out. As a baseline, two layers of molybdenum at the top and bottom of the plastic bars (see Figure \ref{fig:cusp_design}) are added to absorb backscattered photons to prevent them from propagating in the sensitive volume. Possible improvements of the modulation factor and reduction in spurious modulation by adding molybdenum layers is assessed to maximise the polarimetric sensitivity of the polarimeter.

\begin{figure}[H]
  \centering
  \includegraphics[trim=0pt 0pt 0pt 0pt, clip,scale=0.5]{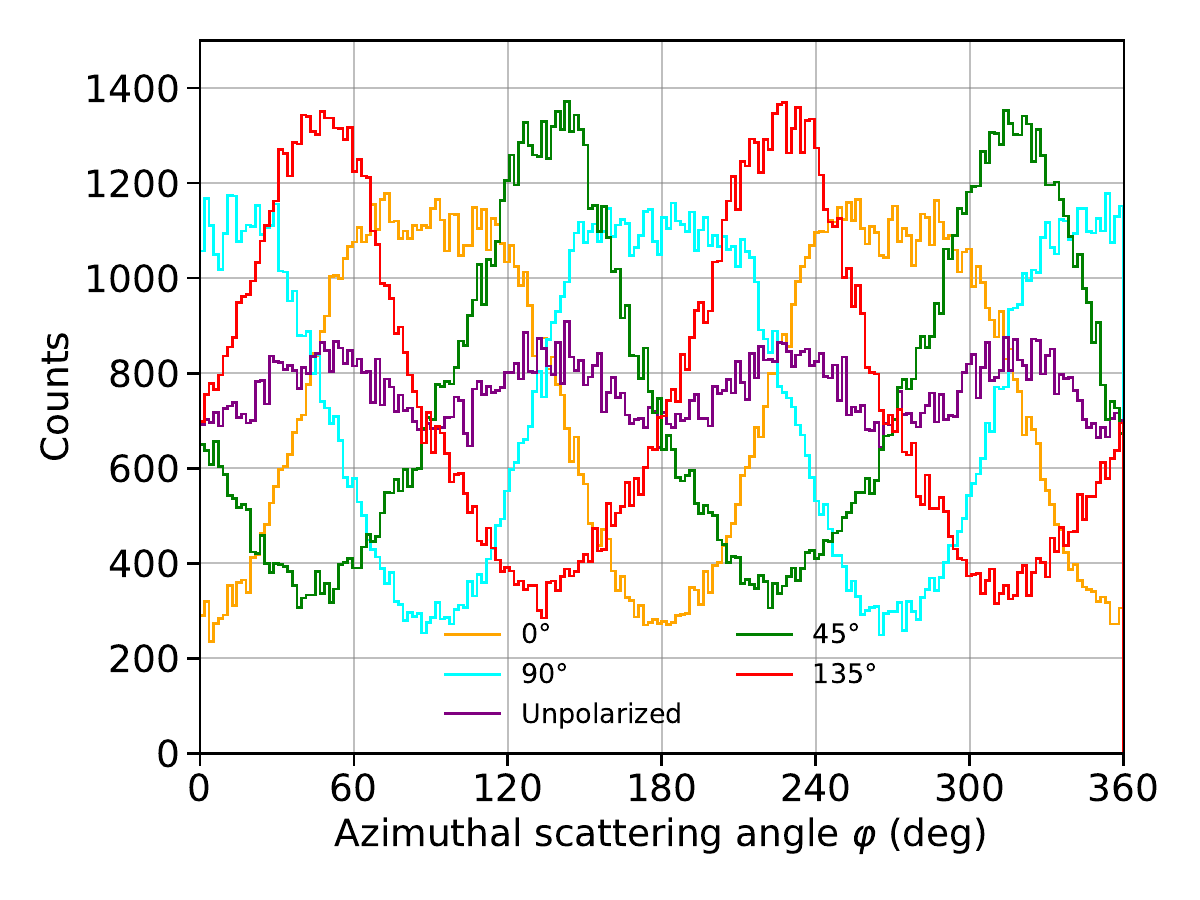} 
  \caption{Modulation induced by geometry for unpolarized radiation shown in violet and modulation induced by 100$\%$ polarised radiation with different polarization angle is shown with different colors.}
  \label{fig:mod_curve}
\end{figure}

The detector response is simulated using homogeneous on-axis polarised and unpolarised beams with an energy of 60~keV at four polarisation angles (0$^{\circ}$, 45$^{\circ}$, 90$^{\circ}$, 135$^{\circ}$) for molybdenum layers of 200~$\mu$m and 400~$\mu$m. The raw azimuthal distribution of polarised photons at different polarisation angles and for unpolarised photons is shown in Figure \ref{fig:mod_curve}. For polarized radiation, the modulation pattern depends on how it combines with the four peaks of spurious modulation (see Figure \ref{fig:mod_curve}), which does not exactly follow the cosine distribution and needs to be decoupled to obtain the polarisation information. We decoupled the spurious modulation from the source using the technique discussed in the previous section.

\subsection{Estimation of modulation factor}

\begin{figure}[H]
  \centering
  \begin{subfigure}[b]{\textwidth}
    \centering
    \includegraphics[width=0.6\textwidth]{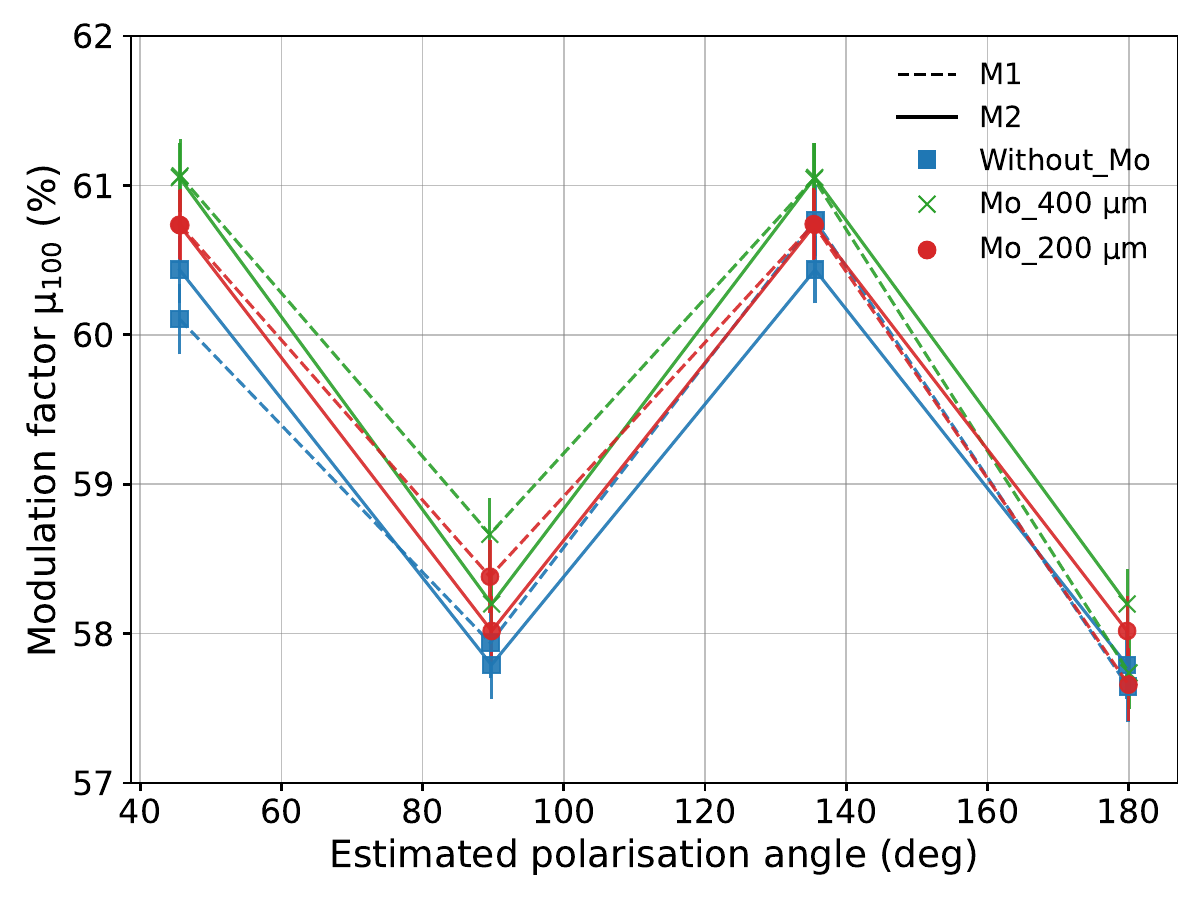}
  \end{subfigure}
  \vspace{0.01\textwidth}
  \begin{subfigure}[b]{\textwidth}
    \centering
    \includegraphics[width=0.62\textwidth]{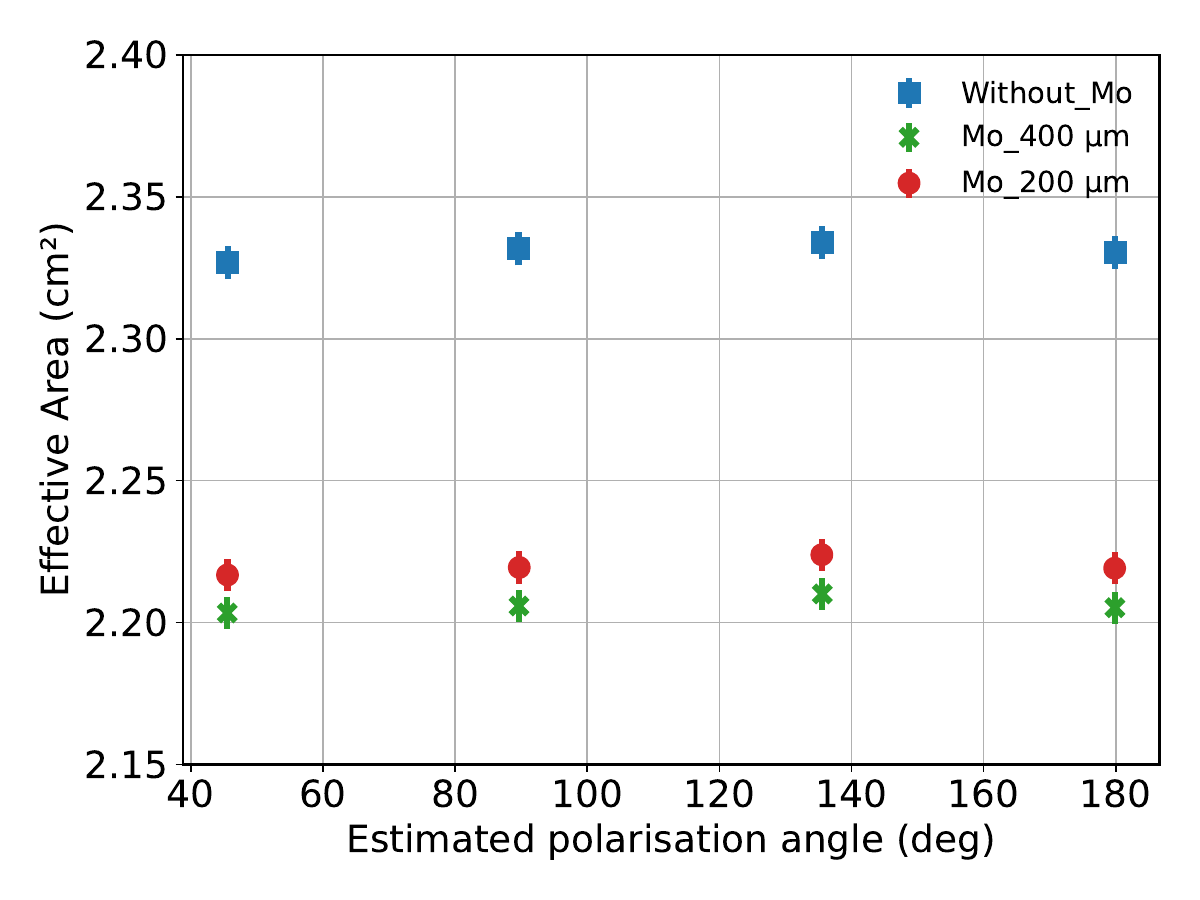}
  
  \end{subfigure}
  \caption{CUSP modulation factor (top) and effective area (bottom) at 60~keV as a function of polarisation angle for 100$\%$ polarised radiation.}
  \label{fig:mod_effective_area_var}
\end{figure}

By computing the Stokes parameters, we derived the modulation factor at 60~keV with two different methods, M1 and M2, as discussed in section \ref{sec:method}. A variation in the modulation factor with polarization angle is observed (see Figure \ref{fig:mod_effective_area_var}), which is a geometrical effect. This arises due to the square configuration of the CUSP detector, which offers better sampling of azimuthal angle along diagonals than edges, resulting in a higher modulation factor at the corner angles. To further validate this, we also checked the effective area as a function of polarisation angle to ensure that the observed variation is not caused by it. It is found that the effective area is constant with angle (see Figure \ref{fig:mod_effective_area_var}), which implies the variation in modulation factor is indeed because of the better sampling at corner angles.

The effect of the molybdenum on the modulation factor and effective area is also studied. Preliminary results indicate that the inclusion of molybdenum improved the modulation factor by approximately. 1.7$\%$ (see Figure \ref{fig:mod_effective_area_var}) in the best possible case, reducing the backscattering from the passive structure of the payload, but it reduces the effective area by 5$\%$ (see Figure \ref{fig:mod_effective_area_var}). However, the final decision on whether to include molybdenum shielding will be based on its combined impact on both the energy spectrum and the modulation factor. The evaluation of its effect on the energy spectrum is currently in progress. Additionally, a discrepancy is noted between the results obtained from the two Stokes analysis methods (M1 and M2). Further investigation is underway to understand the origin of this difference.

\subsection{Spurious modulation due to geometric effects}

\begin{figure}[H]
  \centering
  \includegraphics[trim=0pt 0pt 0pt 0pt, clip,scale=0.5]{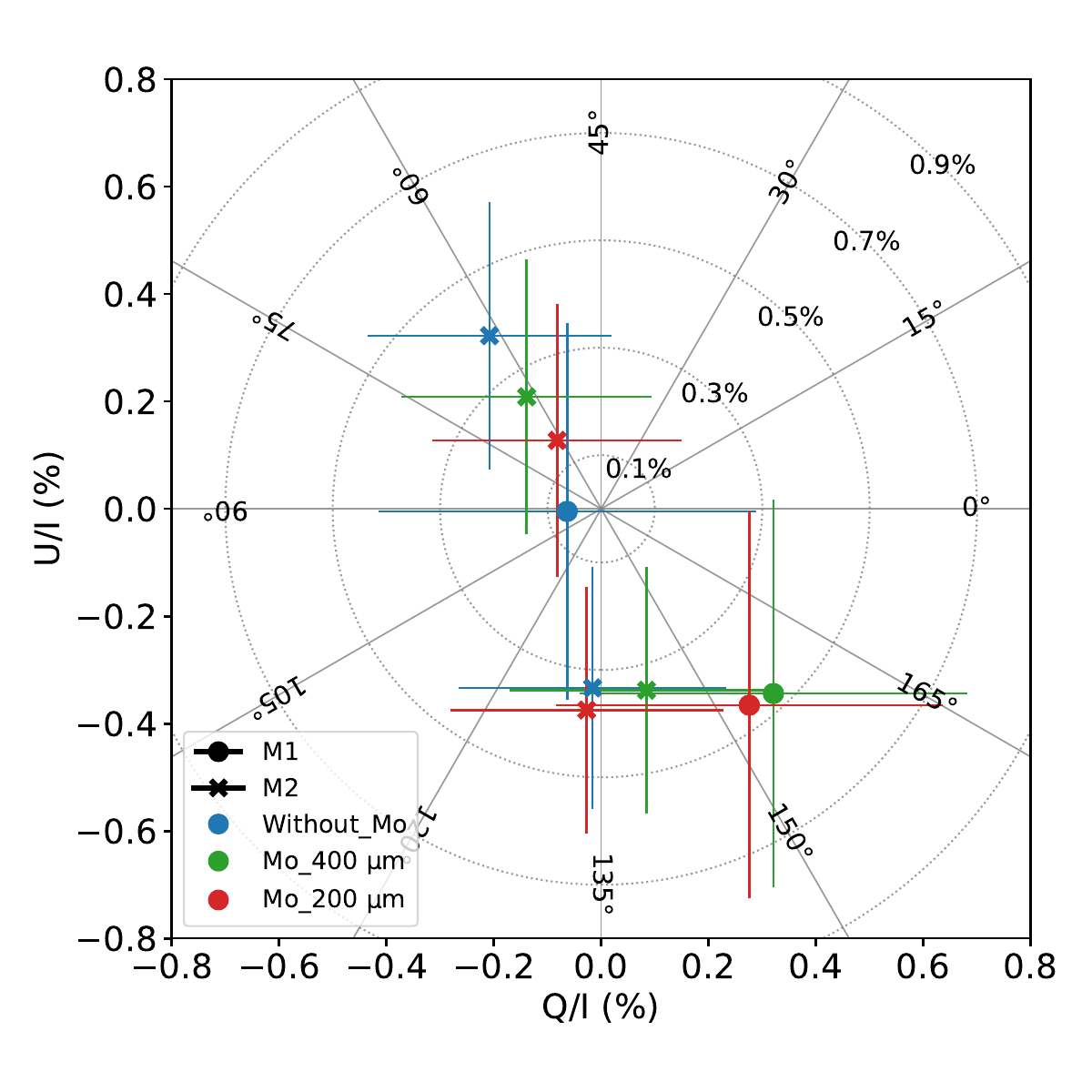} 
  \caption{Normalised q and u for four different polarisation angles using two methods. The diagonal lines indicate the polarization angles, while the concentric circles represent the spurious modulation amplitude.}
  \label{fig:mod_spurious}
\end{figure}

We derived the spurious modulation (sm) at 60~keV with the two independent analysis methods (M1 and M2) to quantify the spurious modulation due to the geometric asymmetry of the instrument.  We evaluated the level of spurious modulation under different backscattering shielding configurations. We compared the scenarios with and without molybdenum shielding to understand the effect of secondary scattering on the spurious modulation factor. In both shielding configurations, the spurious modulation amplitude is found to be below 1$\%$ in the second harmonic (see Figure \ref{fig:mod_spurious}). The normalized q and u parameters obtained using methods M1 and M2 are consistent within 2$\sigma$, even in the worst-case scenario. This level of agreement indicates the absence of any significant method-dependent bias.

\section{Conclusion}

We developed a Monte Carlo code based on Geant4 to evaluate and optimise the scientific performance of the CUSP using a detailed mass model of the instrument and also an analysis method to extract the polarimetric information from the data. This code and data analysis method will be used to evaluate the polarimetric response of the instrument. This will also be a useful tool for calibrations and the development of data analysis software.

Here, we have shown the initial results obtained using the current mass model and explored the Stokes parameter formalism to obtain the polarimetric information. We also carried out the optimisation of the CUSP mass model to mitigate the backscattering events, which introduce the unpolarised component in the results. As a first result, the molybdenum shielding does not improve the modulation factor significantly by stopping the backscattering. Variation in the modulation factor is observed due to geometric effects inherent to the detector design. The spurious modulation in the second harmonic is found to be less than 1$\%$, which suggests a very small contribution of spurious amplitude due to geometric effects.

We aim to evaluate CUSP's scientific performance using more complex analysis methods by estimating the MDP and spurious modulation across different classes of solar flares, accounting for geometric effects and in-orbit background. Further design optimisation will also be carried out.

\acknowledgments 

This work is funded by the Italian Space Agency (ASI) under the Alcor program. Program, within the development of the CUbesat solar polarimeter ( CUSP ) mission under the ASI-INAF contract n. 2023-2-R.0.


\bibliographystyle{spiebib} 

\end{document}